# PARTITION FUNCTION FOR THE TWO-DIMENSIONAL SQUARE LATTICE ISING MODEL IN A NON-ZERO MAGNETIC FIELD- A HEURISTIC ANALYSIS

G.Nandhini and M.V.Sangaranarayanan*

Department of Chemistry

Indian Institute of Technology - Madras, Chennai – 600036

India

Email : sangara@iitm.ac.in

Fax : +9144 22570545

**Abstract**

The exact partition function of the two-dimensional nearest neighbour Ising model pertaining to square lattices is derived for N sites in the case of a non-vanishing magnetic field. When the magnetic field is zero, the partition functions estimated from the present analysis are identical with those arising from Onsager's exact solution.

*1. Introduction*

An exact expression for computing the partition function for *two-dimensional* Ising models has remained elusive for more than six decades [1] although the one-dimensional case is pedagogical [2]. This inability which arises solely due to the dimensionality of the problem is difficult to comprehend. This inadequacy may partially be attributed to the manner in which the one-dimensional Ising model is solved viz. the transfer matrix method [3].Hence it is imperative to investigate other methods of estimating partition functions without involving the transfer matrix method. In this Communication, we report a closed form expression for the partition function of the two-dimensional nearest neighbour Ising model satisfying periodic boundary conditions for any square lattice of N sites.

*2. Methodology*

The Ising Hamiltonian $H_T$ in the case of the external magnetic field H and nearest neighbour interaction energy J may be represented as[3]

$$H_T = -J\sum_{\langle ij\rangle} (\sigma_{i,j}\sigma_{i,j+1} + \sigma_{i,j}\sigma_{i+1,j}) - H\sum_{\langle ij\rangle} \sigma_{i,j} \qquad (1)$$

The canonical partition function pertaining to the above Hamiltonian is

$Q = \text{Tr}\{\exp(-H_T/kT)\}$ (2)

The simplest square lattice for which all the configurations can be enumerated is a square lattice of 16 sites for which the sum of $2^{16}$ configurations yields directly the canonical partition function Q. This is easy to accomplish and yields in turn, estimates of Q for any value of H and J albeit for a *small finite lattice*. However, it is possible to generalize the result obtained using the enumeration strategy in a heuristic manner by two different methods and the following equation for Q is easily deduced for any square lattice of N sites viz.

$Q(J, H) = \{2e^{32J/kT}\cosh(16H/kT) + 32e^{24J/kT}\cosh(14H/kT) + 16e^{16J/kT}(2e^{8J/kT}+13)$
$\cosh(12H/kT) + 32e^{8J/kT}(e^{16J/kT} + 12e^{8J/kT} + 22)\cosh(10H/kT) + 8(4e^{24J/kT} + 66e^{16J/kT} + 220e^{8J/kT}$
$+165)\cosh(8H/kT) + 32(e^{24J/kT} + 20e^{16J/kT} + 90e^{8J/kT} + 120 + 42e^{-8J/kT})\cosh(6H/kT) + 16(2e^{24J/kT} +$
$45e^{16J/kT} + 240e^{8J/kT} + 420 + 252e^{-8J/kT} + 42e^{-16J/kT})\cosh(4H/kT) + 32(e^{24J/kT} + 24e^{16J/kT}$
$+140e^{8J/kT} + 280 + 210e^{-8J/kT} + 56e^{-16J/kT} + 4e^{-24J/kT})\cosh(2H/kT) + [16e^{8J/kT}(e^{16J/kT}+147) +$
$392(e^{16J/kT} + 3e^{-16J/kT}) + 4900 + 2e^{-8J/kT}(e^{-24J/kT} + 56e^{-16J/kT} + 1960)]\}^{N/16}$

(3)

### *3. Results and Discussion*

The above equation constitutes the first ever closed-form expression for Q pertaining to square lattices with nearest neighbour interactions. When H = 0, the above equation reduces to

$Q(J, H=0) = \{4[\cosh(32J/kT) + 120\cosh(24J/kT) + 1820\cosh(16J/kT) + 8008\cosh(8J/kT) +$
$6435]\}^{N/16}$ (4)

In Fig 1, estimates of Q arising from the above equation are compared with those predicted by Onsager's exact solution [4] for various values of N viz. N =16; 64; 144; 256 etc and the two estimates are entirely consistent for all ranges of J. In contrast to the Onsager's exact solution for Q in the case of H= 0 involving a numerical integration, the above equation (4) involves an algebraic computation with an inherent symmetry associated with it.

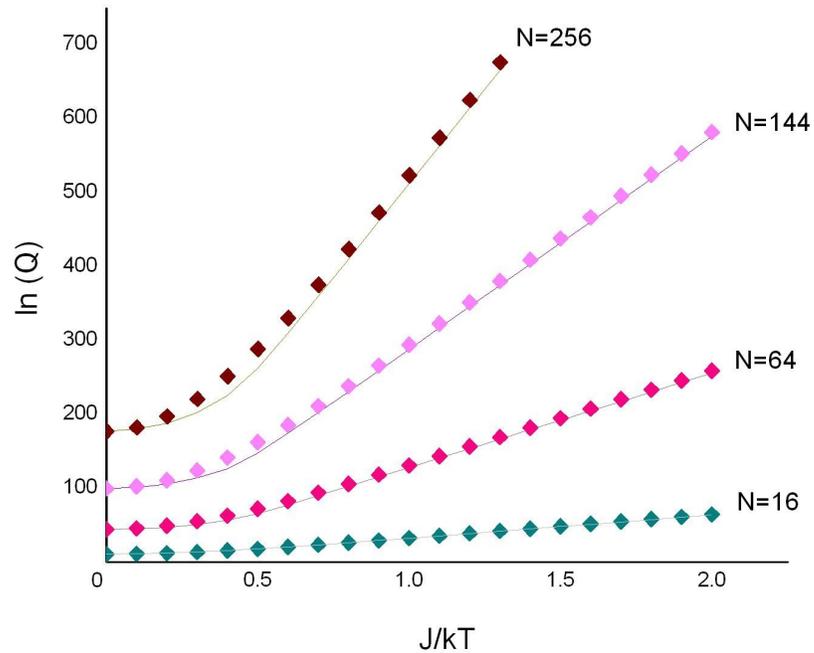

Fig.1: A comparison of the partition function values arising from equation (4) with that of Onsager's analysis [2,4]. Symbols indicate the estimates of equation (4) while the lines represent the values predicted by the Onsager's exact solution.

It is important to reiterate that the above equation has its origin in the enumeration of 65,536 configurations for a square lattice of 16 sites and subsequent extrapolation to general lattice sizes. It appears that the enumeration of configurations is an easier method for deriving the partition function in contrast to the transfer matrix approach or its variants.

The spontaneous magnetization is a crucial parameter [5] for the analysis of two-dimensional phase transitions. A preliminary study using equation (3) indicates that the dimensionless critical temperature ($J/kT_c$) is 0.4418 which is in excellent agreement with the value of Yang and Lee [5] while the corresponding 'dimensionless' magnetic field is $\sim 10^{-2}$ in stead of zero.

## 4. Summary

An exact expression for the partition function of two-dimensional nearest neighbour Ising models for square lattices is derived in a heuristic manner for non-zero magnetic field. The validity of the proposed expression is demonstrated by comparing the partition functions with the values predicted by Onsager's exact solution, when the magnetic field is zero.

**References**
1. Cipra, B., Statistical Physicists Phase out a Dream. *Science 2.* **288,** 1561 – 1562 (2000).
2. Chandler, D., *Introduction to Statistical Mechanics Ch.5* (Oxford University Press, Oxford, 1987).
3. Hill, T. L. *Statistical Mechanics – Principles and Selected Applications* (McGraw-Hill Book Company, USA, 1956).
4. Onsager, L. Crystal Statistics. I. A Two-Dimensional Model with an Order-Disorder Transition. *Phys.Rev.* **65,**117-149 (1944).
5. Lee, T.D. & Yang, C.N. Statistical theory of equations of state and phase transitions. II. Lattice gas and Ising Model. *Phys. Rev.* **87,** 410 (1952).